# Unified HT-CNNs Architecture: Transfer Learning for Segmenting Diverse Brain Tumors in MRI from Gliomas to Pediatric Tumors

Ramy A. Zeineldin[1,2*]  · Franziska Mathis-Ullrich[1]

[1]Department of Artificial Intelligence in Biomedical Engineering (AIBE), Friedrich-Alexander-University Erlangen-Nürnberg (FAU), Germany
[2]Faculty of Electronic Engineering (FEE), Menoufia University, 32952 Menouf, Egypt

**Abstract**

**Purpose** Accurate segmentation of brain tumors from 3D multimodal MRI is vital for diagnosis and treatment planning across diverse brain tumor types. This paper addresses the challenges posed by the BraTS 2023, presenting a unified transfer learning approach that applies to a broader spectrum of brain tumors, including Glioma and Pediatric Tumors.

**Methods** We introduce HT-CNNs, an ensemble of Hybrid Transformers and Convolutional Neural Networks optimized through transfer learning for varied brain tumor segmentation. This method captures spatial and contextual details from 3D MRI data, fine-tuned on diverse datasets representing common tumor types. Through transfer learning, HT-CNNs utilize the learned representations from one task to improve generalization in another, harnessing the power of pre-trained models on large datasets and fine-tuning them on specific tumor types. Specifically, we employed two primary transfer learning techniques: fine-tuning pre-trained models on the BraTS dataset for Adult Glioma segmentation and ensemble learning using the STAPLE method. We preprocess diverse datasets, including patients from multiple international distributions, ensuring representativeness for the most common brain tumors. Our rigorous evaluation employs standardized quantitative metrics across all tumor types, ensuring robustness and generalizability.

**Results** The proposed ensemble model achieves superior segmentation results across the BraTS validation datasets over the previous winning methods. Comprehensive quantitative evaluations using the dice similarity coefficient (DSC) and Hausdorff distance (HD95) demonstrate the effectiveness of our approach. Further, fine-tuning significantly improved the performance of baseline models, with notable increases in DSC and decreases in HD95. For instance, the DSC for Pediatric brain tumors increased from 0.4097 to 0.6248, while the HD95 decreased from 163.37 to 37.46. Similarly, for the Sub-Saharan Africa datasets, the DSC improved from 0.7832 to 0.8647, and the HD95 decreased from 18.38 to 10.98. Qualitative segmentation predictions further validate the high-quality outputs produced by our model.

**Conclusion** Our findings underscore the potential of transfer learning and ensemble approaches in medical image segmentation, indicating a substantial enhancement in clinical decision-making and patient care. Despite facing challenges related to post-processing and domain gaps, our study sets a new precedent for future research aimed at refining brain tumor segmentation models. The docker image for the code and models has been made publicly available (https://hub.docker.com/r/razeineldin/ht-cnns) to encourage further advancements in the field.

**Keywords** Brain Tumor Segmentation. Ensemble Learning . Multimodal MRI . Hybrid Transformer . Transfer Learning.

# Introduction

Brain tumors, particularly glioblastoma (GBM) and diffuse astrocytic glioma with molecular features of GBM (WHO Grade 4 astrocytoma) represent the most aggressive and common malignant primary tumors of the central nervous system in adults [1,2]. These tumors exhibit significant appearance, shape, and histology heterogeneity, making their accurate identification and localization a critical challenge. GBM patients face a grim prognosis, with an average survival of only 14 months under standard care treatment and just four months if left untreated. Despite numerous experimental treatment approaches proposed over the past two decades, there have been limited improvements in patient prognosis. Brain tumors' lethality and the complexities in their detection and treatment underscore the urgent need for more efficient and accurate automated segmentation solutions to assist clinicians in managing these conditions.

The segmentation of gliomas, encompassing distinct sub-regions such as the enhancing tumor (ET), peritumoral edematous/invaded tissue (ED), and the necrotic components of the core tumor (NCR), presents a formidable challenge in the medical field. While manual segmentation serves as the gold standard for various clinical applications, including neurosurgical planning, image-guided interventions, and tumor growth monitoring, this approach is labor-intensive, subjective, and highly reliant on the expertise of clinicians. This manual process becomes impractical when dealing with numerous patients, highlighting the critical need for automated deterministic segmentation solutions to streamline and expedite this crucial task. Accurately identifying boundaries for brain tumor sub-regions in MRI scans is essential for improving surgical treatment planning, generating radiotherapy maps, and facilitating follow-up procedures, ultimately enhancing patient care in these formidable malignancies.

The Medical Image Computing and Computer-Assisted Interventions Brain Tumor Segmentation Challenge (MICCAI BraTS) is a significant initiative dedicated to addressing the challenging problem of the automated brain tumor sub-region segmentation [3,4]. Established in 2012, this challenge has played a pivotal role in advancing machine learning (ML) in glioma diagnosis. Each year, BraTS provides a comprehensive dataset of multi-parametric MRI scans and ground-truth annotations, offering a benchmark for developing and evaluating state-of-the-art algorithms in tumor segmentation, classification, and, more recently, survival prediction [2,5,6]. The BraTS challenges have been pivotal in fostering collaboration among researchers, clinicians, and institutions to enhance the accuracy and efficiency of brain tumor diagnostics. This ongoing effort aims to improve automated brain tumor segmentation methods and contribute to the overall progress in glioma diagnostics and treatment planning.

In the latest edition of the BraTS challenge, BraTS 2023, a cluster of challenges has been introduced to address various crucial aspects of brain tumor diagnostics and treatment [7,8]. These new challenges include sub-Saharan African patient populations (SSA), meningioma segmentation (MEN), brain metastasis segmentation (MET), pediatric brain tumors (PED), global and local missing data, augmentation techniques, and algorithmic generalizability in addition to the original adults' glioma segmentation (GLA). By expanding the focus to incorporate these diverse aspects, the BraTS 2023 challenge aims to create a comprehensive benchmarking environment, further advancing the field and facilitating the development of innovative algorithms. Additionally, introducing a specialized pediatric brain tumor segmentation challenge underscores the pressing need to improve diagnostics and treatment planning for this vulnerable patient group. The BraTS challenges continue to be instrumental in pushing the boundaries of medical image analysis, contributing to advancing brain tumor research and clinical care.

Automated deep learning methods have the potential to revolutionize the brain tumor segmentation process, significantly enhancing accuracy, efficiency, and clinical utility. Convolutional Neural Networks (CNNs) have been particularly effective in medical image analysis, including the segmentation of brain tumors [9,10]. CNNs excel at learning intricate features from complex data, making them well-suited for capturing the nuanced and diverse characteristics of brain tumors evident in multi-parametric MRI scans. By training on large annotated datasets, CNNs, specifically U-Net and its numerous variants [11,12], can learn to differentiate tumor sub-regions, such as the enhancing tumor, peritumoral edema, and necrotic components, providing precise and consistent segmentation. The architecture of U-Net, with its encoder for feature extraction, decoder for spatial detail recovery, and skip connections for feature fusion, has been fundamental in this field. The upper layers of U-Net, capturing broad contextual information, and lower layers, rich in spatial detail, work in concert to enhance segmentation results. Variants like NAG-Net [13] and SBANet [14] have pushed the envelope by integrating clinical domain knowledge and segmentation-based attention to refine feature extraction and tumor recognition capabilities. However, challenges persist, especially when dealing with images that have complex backgrounds, irrelevant noise, and unclear boundaries, which are common in medical imaging.

Parallel to the U-Net evolution, Transformer-based methods [15] have been carving out a significant role in medical image analysis. Transformers, originally developed for natural language processing, offer exciting prospects for brain tumor segmentation. Transformers employ self-attention mechanisms to capture long-range dependencies, making them particularly effective at understanding the spatial relationships and contextual information crucial for accurate tumor segmentation. Models like MESTrans [16] and TransXAI [17] have shown that Transformers can indeed excel in this domain. Nevertheless, the high computational cost associated with the self-attention operation has limited their practicality, especially in cases with limited datasets. In response, recent efforts have been directed toward streamlining the self-attention mechanism. CASF-Net [18] and UTNet [19] are prime examples, with the former utilizing a factorized attention mechanism and the latter projecting features into a lower-dimensional space to alleviate computational demands. Despite achieving impressive performance, there remains the risk of neglecting small yet clinically significant features due to these reductions in dimensionality and complexity.

These two strands of research lay the foundation for our proposed HT-CNNs framework. The combination of CNNs and Transformers can further amplify the effectiveness of automated brain tumor segmentation. By leveraging the strengths of both architectures, deep learning models can comprehensively analyze multi-parametric MRI data, extracting intricate spatial and contextual features while maintaining fine-grained accuracy. These advanced methods, trained on large and diverse datasets, can streamline the segmentation process, reducing the reliance on manual labor, minimizing subjectivity, and expediting the diagnosis and treatment planning for brain tumors. By merging these two powerful architectures and employing transfer learning, we strive to create a segmentation tool that is not only accurate and efficient but also robust against the variability and complexity inherent in medical imaging tasks. As the field of deep learning continues to evolve, it holds the promise of contributing significantly to improving patient care and outcomes in brain tumor management.

In this work, we present a unified framework that integrates Hybrid Transformers with CNNs, utilizing transfer learning to bridge the gap between various brain tumor types and imaging modalities. Our contributions are as follows: We propose a novel HT-CNNs architecture that combines the local feature extraction capabilities of CNNs with the global contextual understanding of Transformers, tailored for the segmentation of Glioma, Intracranial Meningioma, Pediatric Tumors, and Brain Metastasis. We leverage transfer learning to enhance model performance across diverse datasets, including those from multiple international consortia, ensuring our approach's applicability to real-world clinical scenarios. In addition, a comprehensive two-stage post-processing strategy was introduced to refine the tumor boundary predictions. The first stage replaces enhancing tumor predictions below a threshold with necrosis, while the second stage employs connected component analysis to handle small enhancing tumor regions. Furthermore, we introduce a rigorous evaluation protocol using standardized quantitative metrics to ensure the robustness and generalizability of our model across all tumor types. This approach not only sets a new benchmark for brain tumor segmentation but also advances the path toward clinical translation, offering a robust tool for assisting clinicians in the challenging task of tumor delineation.

## Methods

### Data

The data used in this paper originates primarily from the BraTS 2023 cluster of challenges, which encompasses a diverse range of datasets representing various brain tumor scenarios. The BraTS dataset consists of retrospective multi-parametric MRI scans collected under standard clinical conditions from multiple institutions with different equipment and imaging protocols, reflecting the broad clinical practice diversity [8,7,2]. Ground truth annotations for tumor sub-regions, including the tumor core (TC), enhancing tumor (ET), peritumoral edema (ED), and necrotic and non-enhancing tumor core (NCR/NET), were meticulously reviewed by expert neuroradiologists.

For specific challenges within BraTS 2023, such as the Sub-Saharan African (SSA) populations, the SSA dataset is utilized, providing annotated pre-operative glioma data from various institutions and scanners and maintaining real-world heterogeneity [7]. These datasets are essential for developing and evaluating automated brain tumor segmentation methods, enabling the creation of robust and generalizable models for improved clinical applications.

Following the established practice of algorithmic assessment in machine learning, the data utilized for the BraTS 2023 challenges has been partitioned into distinct sets: a training set, which accounts for 70% of the data; a validation set, representing 10%; and a separate testing dataset, constituting 20% of the total data.

### HT-CNNs Architecture

**Overall Architecture.** The HT-CNNs framework introduces a comprehensive architecture that integrates multiple advanced neural network models to address the complexity of brain tumor segmentation (Fig. 1). At the forefront, the framework accepts an input MRI, which undergoes an initial preprocessing block designed to standardize and enhance the imaging data for subsequent analysis. Following preprocessing, the data is fed into a tripartite network system comprising:

1. **nnU-Net (CNN)**: This component serves as the convolutional neural network, focusing on extracting spatial hierarchies and detailed textural information from the preprocessed MRI data.
2. **TransBTS (Transformer)**: The TransBTS network employs a Transformer architecture, adept at capturing long-range dependencies and global contextual relationships inherent in the imaging data.
3. **DeepSCAN (Attention)**: As the attention-focused component, DeepSCAN processes the feature maps from the previous networks, applying attention mechanisms to refine the feature representation further and emphasize salient tumor regions.

Each network includes a dedicated local post-processing block, fine-tuning the network-specific outputs to ensure high-quality segmentation maps. These refined outputs are then synergistically combined using the Simultaneous Truth and Performance Level Estimation (STAPLE) method [20]. This ensemble technique effectively integrates the strengths of the individual models, yielding a robust final output segmentation that embodies both precise local details and comprehensive global context, essential for accurate tumor delineation.

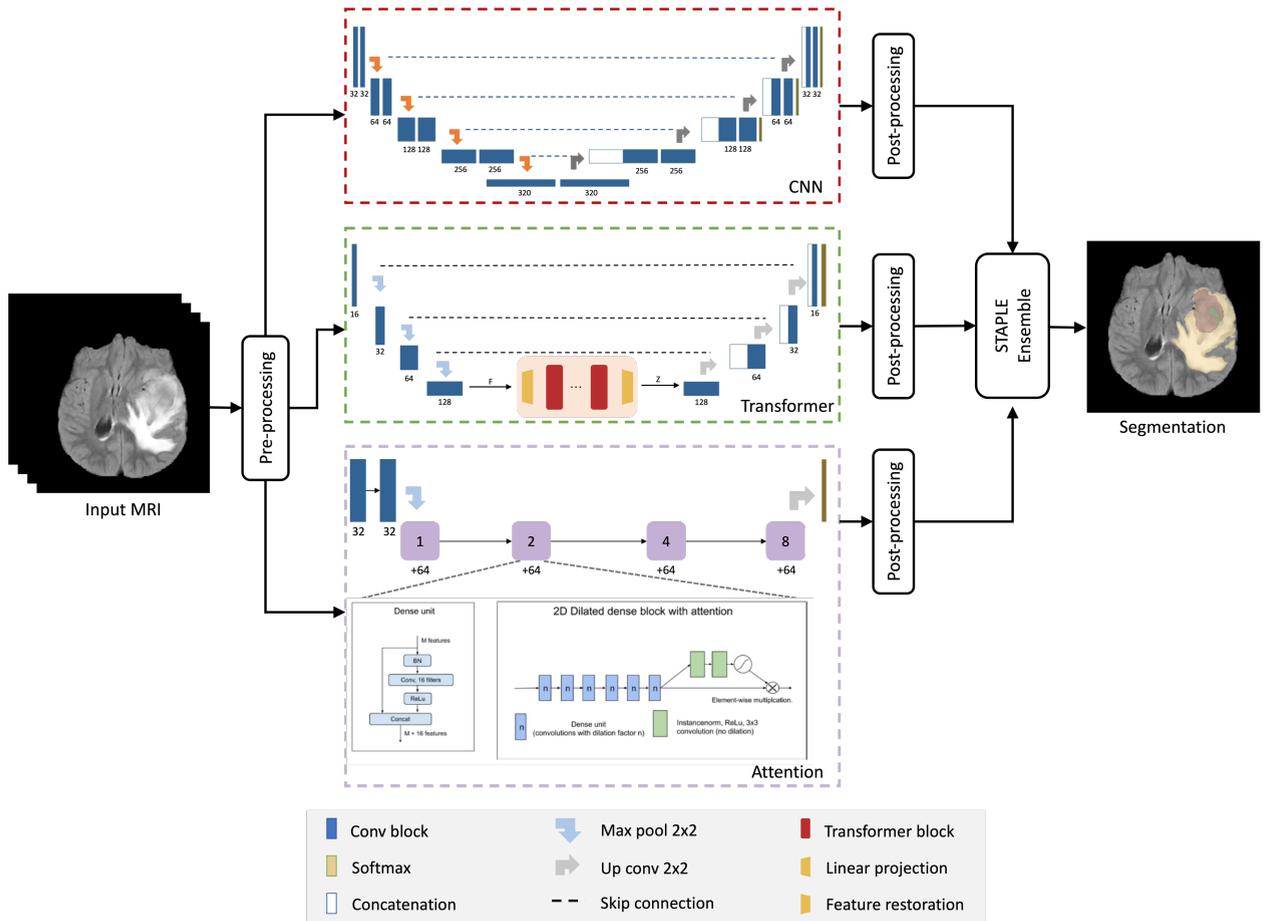

**Fig. 1** HT-CNNs Architecture for Brain Tumor Segmentation.

**Baseline U-Net.** The baseline network architecture is based on the widely-used 3D U-Net encoder-decoder structure [21,10,11], which has been proven effective in various medical image segmentation tasks, as shown in Fig. 1. The network consists of two main components: the encoder, which captures high-level features from the input data, and the decoder, which upsamples these features to produce the final segmentation map. The encoder part of the network consists of multiple consecutive 3D convolutional blocks. Each block contains two 3 × 3 × 3 convolutional layers, followed by a Rectified Linear Unit (ReLU) activation function. Max-pooling with a kernel size of 2 × 2 × 2 for downsampling is employed, reducing the spatial dimensions of the feature maps while preserving important information. The number of filters in the convolutional layers increases progressively through the encoder, allowing the network to capture increasingly complex features. In the decoder part, the upsampled feature maps from the encoder are passed through 3D transposed convolutional layers, also known as deconvolutional layers. These layers restore the spatial resolution of the feature maps, enabling the network to produce a detailed segmentation map. Skip connections interconnect the encoder and decoder, allowing the network to integrate high-resolution information from the encoder with the upsampled features from the decoder. Finally, a 1 × 1 × 1 convolutional layer with softmax activation generates the probability distribution of the tumor regions in the input volume.

**Transformer Network**. To handle the computational complexity of the Transformer for 3D volumetric data, we adopt a method inspired by the Vision Transformer (ViT) [22] (as detailed in ref [23]), which splits the image into fixed-size (16×16) patches and reshapes each patch into a token. This approach effectively reduces the sequence length, making it computationally tractable. However, for volumetric data, directly applying this tokenization strategy can hinder the model's ability to capture local context information across spatial and depth dimensions essential for volumetric segmentation. To address this challenge, they employed a solution that involves stacking 3 × 3 × 3 convolution blocks with downsampling (strided convolution with stride=2). This transformation encodes the input images into a low-resolution/high-level feature representation $F \in \mathbb{R}^{K \times \frac{H}{8} \times \frac{W}{8} \times \frac{D}{8}}$ (K=128), equivalent to 1/8 of the input dimensions of H, W, and D (overall stride OS=8). The enriched local 3D context features within F enhance the model's capability to analyze volumetric data. This feature map F is then fed into the Transformer encoder to learn long-range correlations further and establish a global receptive field. This approach allows the model to effectively handle volumetric data, considering both local and global contextual information, thereby elevating its segmentation performance. The specific architectural and transformational details of the feature embedding and Transformer encoder can be found in the corresponding sections of ref [23].

**Attention Network.** To further improve the baseline 3D U-Net architecture, we incorporated an axial attention decoder, a critical enhancement that leverages the power of self-attention mechanisms. Self-attention, originally introduced and popularized in natural language processing [15], has gradually made its way into the computer vision domain [22]. However, applying self-attention to vision problems, especially with 3D data, has been challenging due to its quadratic computational complexity with input size. To address this, axial attention was introduced as an efficient solution for multi-dimensional data [24]. By independently applying self-attention to each axis, the computational complexity scales linearly with image size, making integrating the attention mechanism with 3D data feasible. In our enhancement, axial attention was integrated into the decoder of the network, specifically applied to the output of the transposed convolution upsampling, and subsequently summed. The architecture and implementation details of the axial attention decoder are illustrated in Figure 2.

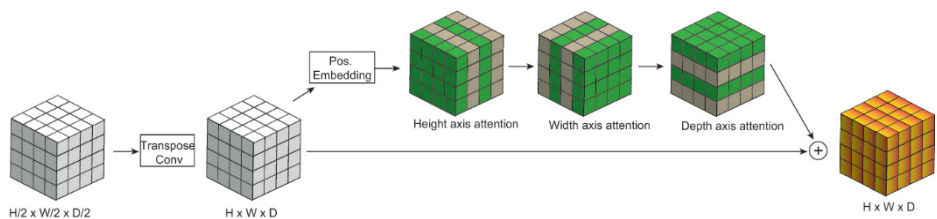

**Fig. 2.** Axial attention is employed individually for each axis in the output obtained from transpose convolution [25].

**Transfer Learning**

In our HT-CNNs framework, transfer learning played a pivotal role in enhancing model performance across diverse brain tumor segmentation tasks. Transfer learning is a machine learning technique where a model developed for a particular task is repurposed as the starting point for a model on a second task [26,27]. It is particularly valuable when available training data is limited or when rapid development is desired. This method has proven to be effective in achieving higher accuracy and faster convergence, especially when pre-trained on a model within a similar domain. Specifically, we utilized two primary techniques for transfer learning:

**Fine-tuning.** In our HT-CNNs framework, we employed fine-tuning as a primary technique for transfer learning. Initially, models were pre-trained on Adult Glioma segmentation tasks using the extensive BraTS dataset (1251 training cases). These pre-trained models provided a rich set of learned features relevant to brain tumor segmentation. For the BraTS 2023 challenges, we fine-tuned these models on smaller, specific datasets, namely the Sub-Saharan Africa (SSA) populations (60 training cases) and Pediatric (PED) brain tumors (99 training cases). This process involved adjusting the pre-trained model parameters to better fit the new datasets, thereby enhancing segmentation accuracy and generalization to different brain tumor types and demographics.

**Ensemble learning.** The second technique employed was ensemble learning using the STAPLE method [20] for transfer learning. This approach integrated the outputs of multiple models to create a consensus segmentation. By leveraging the strengths of various models, the STAPLE technique helped mitigate individual model biases and improved overall segmentation robustness. This ensemble method was particularly effective in refining the segmentation results and ensuring high-quality outputs across the diverse datasets of the BraTS 2023 challenges.

**Post-processing Strategy**

Our post-processing strategy addresses several BraTS-specific challenges by addressing two scenarios to refine the predictions. Firstly, in cases where the reference segmentation lacked enhancing tumor regions, we removed enhancing tumor predictions below a certain threshold, replacing them with necrosis predictions. This approach optimized the model's ranking while managing potential true positive losses, with the threshold fine-tuned via cross-validation on the mean Dice and ranking scores. Additionally, we employed connected component analysis to handle small enhancing tumor regions. Components smaller than 16 voxels with a mean probability less than 0.9 were classified as necrosis, ensuring consistency. Moreover, if the predicted enhancing tumor voxel count fell below 73 voxels with a mean probability less than 0.9, we replaced all enhancing tumor voxels with necrosis predictions. This post-processing strategy accounted for scenarios where enhancing tumor regions were limited or absent, resulting in a more balanced evaluation and enhancing the overall robustness of our model.

## Experiments and Results

**Experimental Setup**

In order to evaluate the performance of our HT-CNNs model, we employed both classic and lesion-wise evaluation metrics. The classic metrics, including the Dice Similarity Coefficient (DSC) and Hausdorff Distance 95th Percentile (HD95), were employed to measure the overlap and boundary accuracy between the predicted and ground truth segmentation masks. These traditional metrics provide a standard framework for assessing segmentation quality and facilitate direct comparison with previous state-of-the-art (SOTA) methods.

The Dice Similarity Coefficient (DSC) measures the overlap between the predicted and ground truth segmentation masks. It is defined as:

$$DSC = \frac{2 \times |A \cap B|}{|A| + |B|} \tag{1}$$

where $A$ is the set of predicted tumor voxels and $B$ is the set of ground truth tumor voxels.

The Hausdorff distance (95%) (HD95) measures the spatial distance between the predicted and ground truth boundaries, focusing on the 95th percentile of distances. Mathematically, the HD95 is calculated as follows:

$$HD95 = max\{d_{95}(A,B), d_{95}(B,A)\} \tag{2}$$

where $d_{95}(A, B)$ is the 95th percentile of distances from points in $A$ to the nearest point in $B$.

In addition to the classic metrics, we incorporated the new lesion-wise metrics introduced in the BraTS 2023 challenges [8,7,2], which offer a more granular evaluation of segmentation performance. The lesion-wise Dice Score and lesion-wise HD95 metrics focus on the precision of delineating individual lesions, penalizing missed lesions and false positives.

The lesion-wise DSC evaluates the segmentation quality at the lesion level, penalizing missed lesions and false positives as defined in Equation (3).

$$DSC_{lw} = \frac{\sum_{i=1}^{L} DSC(l_i)}{TP + FN + FP} \qquad (3)$$

where $DSC(l_i)$ is the dice score for each individual lesion $l_i$, and $TP$, $FN$, and $FP$ are the true positives, false negatives, and false positives respectively. $L$ represents the number of tumor regions.

Similarly, the lesion-wise HD95 provides a detailed measure of boundary accuracy for each lesion, penalizing inaccuracies as follows:

$$HD95_{lw} = \frac{\sum_{i=1}^{L} HD95(l_i)}{TP + FN + FP} \qquad (4)$$

where $HD95(l_i)$ is the 95th percentile Hausdorff distance for each lesion $l_i$.

**Implementation Details**

We began our implementation by preprocessing the BraTS data, adapting it for consistent input. This involved essential steps outlined in the BraTS challenge guidelines, such as DICOM to NIFTI format conversion, coordinate system reorientation, 1 × 1 × 1 mm resampling, and skull-stripping. In addition, we cropped brain pixels and resized the resultant images to a spatial resolution of 192 × 224 × 160, achieving a closer field of view (FOV) for efficient deep learning model training. Z-score normalization was applied individually to each MRI image. Our implementation utilized Python with the PyTorch and MONAI libraries, running on an NVIDIA RTX 4090 GPU. We employed a combined Dice and CE loss function for training, optimizing with the Adam optimizer with a cosine annealing learning rate strategy for 500 epochs. The Transformer architecture was used for the experiments, pre-trained on ImageNet-1K weights, with on-the-fly data augmentation techniques, including random flips, intensity shifts, and scaling. During the test phase, the sliding window method with an overlap rate of 0.6 was employed. These streamlined implementation details ensure the reliability and effectiveness of our approach for precise brain tumor segmentation.

To empirically evaluate the efficacy of transfer learning methodologies, we undertook a comprehensive series of experiments utilizing the BraTS 2023 datasets. The experiments involved:

- Evaluating the impact of ensemble learning on segmentation accuracy.
- Comparing the performance of models with and without fine-tuning.
- Analyzing the performance improvements across different tumors.

**Impact of Ensemble Learning**

The performance of our implemented models was evaluated on the validation sets of the BraTS 2023 challenges, and the results are listed in Tables 1. Synapse, the online evaluation platform by Sage Bionetworks, was utilized for the evaluation, leveraging two essential metrics: DSC and the HD95. Our analysis averaged the DSC scores and HD95 values over the three assessed tumor sub-regions, enabling a comprehensive evaluation. We adopted an ensemble approach to achieve a robust evaluation, averaging the results from five distinct models generated using different

8cross-validation training configurations. The summary of these metrics is presented in the final column, providing a comparative ranking of our methods. Furthermore, HT-CNNs model showcases competitive performance in brain tumor segmentation on the GLA validation set, as listed in Table 2. It achieved an average DSC of 0.8842, which is on par with the winner of 2023 and higher than the winners of 2020, 2021, and 2022. Additionally, the HT-CNNs model demonstrates an improvement in the HD95 metric over the 2021, 2022, and 2023 winners, indicating its efficacy in precise tumor delineation.

**Table 1** Performance of five-fold cross-validation models on BraTS 2023 GLA Validation Cases.

| Model | DSC ↑ | | | | HD95 ↓ | | | |
|---|---|---|---|---|---|---|---|---|
| | ET | TC | WT | Avg | ET | TC | WT | Avg |
| nnU-Net | <u>0.8402</u> | <u>0.8718</u> | 0.9213 | <u>0.8778</u> | 16.24 | 9.36 | <u>4.69</u> | 10.09 |
| TransBTS | 0.8303 | 0.8649 | 0.9157 | 0.8703 | 18.33 | 9.24 | 7.17 | 11.58 |
| DeepSCAN | 0.8306 | 0.8683 | <u>0.9228</u> | 0.8739 | **14.86** | <u>8.42</u> | 4.78 | **9.35** |
| HT-CNNs | **0.8449** | **0.8786** | **0.9291** | **0.8842** | <u>16.19</u> | **7.88** | **4.21** | <u>9.43</u> |

- **Bold** and <u>underlined</u> values correspond to best and second-best scores

**Table 2** Comparison of our segmentation model and SOTA methods on the validation set for the GLA.

| Model | DSC ↑ | | | | HD95 ↓ | | | |
|---|---|---|---|---|---|---|---|---|
| | ET | TC | WT | Avg | ET | TC | WT | Avg |
| Winner 2020 [10] | 0.8402 | 0.8718 | 0.9213 | 0.8778 | **16.028** | 8.953 | 3.823 | 9.601 |
| Winner 2021 [25] | <u>0.8451</u> | <u>0.8781</u> | 0.9275 | <u>0.8836</u> | 20.73 | <u>7.623</u> | **3.47** | 10.608 |
| Winner 2022 [28] | 0.8438 | 0.8753 | 0.9271 | 0.8821 | 17.504 | **7.533** | <u>3.595</u> | **9.544** |
| Winner 2023 [29] | **0.8464** | 0.8769 | **0.9294** | **0.8842** | 17.81 | 11.12 | 4.26 | 11.06 |
| HT-CNNs | 0.8449 | **0.8786** | <u>0.9291</u> | **0.8842** | <u>16.189</u> | 7.88 | 4.21 | <u>10.199</u> |

- **Bold** and <u>underlined</u> values correspond to best and second-best scores

**Fine-tuning Results Studies**

To validate the effectiveness of our transfer learning techniques, we conducted a series of comparative studies. These studies compared the performance of models before and after applying transfer learning and ensemble techniques. Standard evaluation metrics, typically DSC and HD95 were used to evaluate segmentation quality. Models pre-trained on Adult Glioma segmentation tasks were fine-tuned on the PED and SSA datasets, showing marked improvements in segmentation performance. The results in Table 3 and Table 4 indicated higher DSC and lower HD95 metrics compared to models trained from scratch, underscoring the benefits of our transfer learning approach.

**Table 3** Fine-tuning performance comparison of HT-CNNs on the BraTS 2023 PED validation cases.

| Model | DSC ↑ | | | | HD95 ↓ | | | |
|---|---|---|---|---|---|---|---|---|
| | ET | TC | WT | Avg | ET | TC | WT | Avg |
| Baseline | 0.1000 | 0.5559 | 0.5733 | 0.4097 | 336.60 | 75.33 | 78.19 | 163.37 |
| Baseline + TR | 0.4026 | 0.6968 | 0.7751 | 0.6248 | 72.44 | 19.12 | 20.83 | 37.46 |
| Baseline + TR + EN | 0.6490 | 0.6165 | 0.9077 | 0.7621 | 60.00 | 13.29 | 3.86 | 25.72 |

- **TR** and **EN** indicate transfer learning and ensemble techniques.

Table 4 Fine-tuning performance comparison of HT-CNNs on the BraTS 2023 SSA validation cases.

| Model | DSC ↑ | | | | HD95 ↓ | | | |
|---|---|---|---|---|---|---|---|---|
| | ET | TC | WT | Avg | ET | TC | WT | Avg |
| Baseline | 0.7064 | 0.8014 | 0.8417 | 0.7832 | 19.02 | 18.17 | 17.96 | 18.38 |
| Baseline + TR | 0.8130 | 0.8575 | 0.9235 | 0.8647 | 11.25 | 10.66 | 11.03 | 10.98 |
| Baseline + TR + EN | 0.8408 | 0.8139 | 0.9604 | 0.8872 | 4.00 | 6.29 | 2.57 | 4.29 |

- **TR** and **EN** indicate transfer learning and ensemble techniques.

The results demonstrate that fine-tuning significantly improves the performance of the baseline models. For the PED validation cases, the DSC increased from 0.4097 to 0.6248, and further to 0.7621 with the addition of ensemble techniques, while the HD95 decreased from 163.37 to 37.46 and further to 25.72, indicating more accurate and reliable segmentation. Similarly, for the SSA validation cases, the DSC improved from 0.7832 to 0.8647 and further to 0.8872 with ensemble techniques, while the HD95 decreased from 18.38 to 10.98 and further to 4.29. These results highlight the importance of transfer learning and ensemble techniques in adapting models to different datasets and enhancing their performance.

**Lesion-wise Metrics Evaluation**

To align with the BraTS 2023 challenge's evaluation standards, we have included lesion-wise metrics in our analysis. These metrics offer a more granular evaluation of segmentation quality, focusing on the precision of delineating individual lesions. The results, depicted in Table 5, illustrate the significant improvements achieved through transfer learning, underscoring the enhanced precision and reliability of our HT-CNNs model in accurately segmenting individual lesions. By incorporating lesion-wise metrics, we provide a comprehensive evaluation that aligns with the latest standards in the field, ensuring our findings are relevant and comparable to contemporary research.

Table 5 Lesion-wise performance comparison of HT-CNNs on the BraTS 2023 validation cases.

| Dataset | Lesion-wise DSC ↑ | | | | Lesion-wise HD95 ↓ | | | |
|---|---|---|---|---|---|---|---|---|
| | ET | TC | WT | Avg | ET | TC | WT | Avg |
| GLA | 0.8187 | 0.8563 | 0.8501 | 0.8417 | 31.82 | 19.08 | 35.84 | 28.91 |
| PED | 0.7023 | 0.6473 | 0.8876 | 0.7675 | 54.24 | 27.41 | 13.11 | 31.58 |
| SSA | 0.6586 | 0.7186 | 0.8596 | 0.7891 | 88.82 | 47.42 | 41.20 | 59.15 |

The winning team of BraTS-PED 2023, CNMCPMI2023, achieved a lesion-wise DSC of 0.65, 0.81, and 0.83 for the ET, TC, and WT regions, respectively [30]. Please note that the winner used an ensemble of nnU-Net and SwinUNETR, which is our second method used for transfer learning. Notably, our HT-CNNs model demonstrates competitive performance, particularly in the WT region, which is crucial for comprehensive tumor segmentation.

**Segmentation Output**

To visually assess the segmentation performance of our HT-CNNs model, we present qualitative segmentation predictions on the BraTS 2023 validation dataset in Figure 3. These results showcase the effectiveness of our approach in accurately segmenting brain tumors across different tasks. The predictions displayed in the figure are generated using our proposed model, and the rows correspond to samples from the GLA, SSA, and PED datasets.

From Figure 3, it is evident that HT-CNNs model produces segmentation results of high quality across various tumor sub-regions and specific tasks. In the GLA dataset, the model successfully identifies and segments the tumor regions, highlighting its capability to precisely delineate the tumor boundaries. Similarly, in the SSA dataset, the model accurately captures the intricate structures of the subventricular/subependymal areas, reflecting its proficiency in segmenting complex brain structures. Moreover, the PED dataset showcases the model's ability to handle pediatric brain tumor segmentation, emphasizing its versatility in tackling diverse brain tumor types.

The qualitative output shown in Figure 3 affirms that HT-CNNs consistently delivers accurate and visually pleasing segmentations across different datasets. The high-quality results validate the effectiveness of our approach in addressing the challenges posed by the BraTS 2023 challenges, further reinforcing the promising performance demonstrated in the quantitative evaluation.

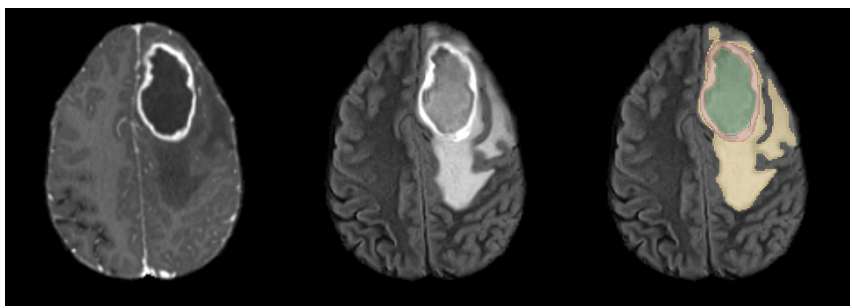

**(a) GLA:** BraTS-GLI-00190-000, EC (0.9848), TC (0.9962), WT (0.9928)

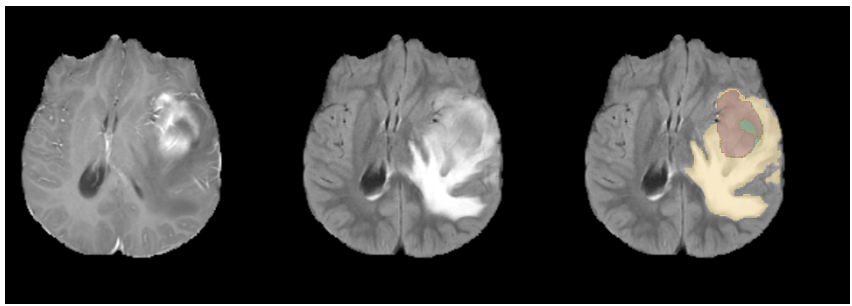

**(b) SSA:** BraTS-SSA-00188-000, EC (0.9196), TC (0.8865), WT (0.9577)

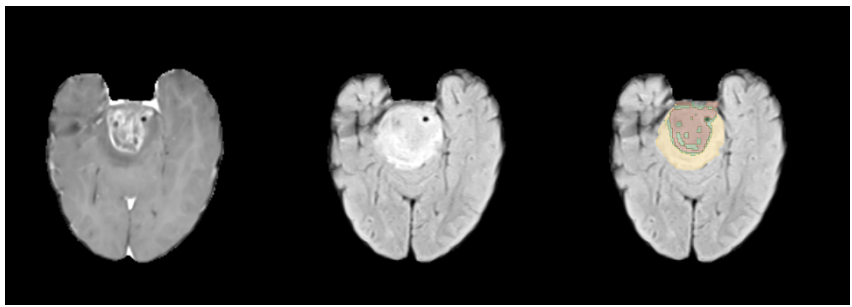

**(c) PED:** BraTS-PED-00030-000, EC (0.9403), TC (0.8823), WT (0.9574)

**Fig. 3** Sample segmentation predictions on the BraTS 2023 validation dataset. Segmentation outcomes generated by our ensemble model are displayed for samples from the GLA, SSA, and PED datasets.

## Discussion

The HT-CNNs framework demonstrates promising advances in the field of medical image segmentation, integrating the precision of U-Net, the contextual awareness of Transformers, and the focus of attention mechanisms. Applied to the BraTS 2023 challenge, our model displayed robust performance, particularly with Glioma and Pediatric Tumors, achieving superior average DSC and competitive HD95 metrics compared to SOTA methods. These outcomes illustrate the model's proficiency in handling the intricacies of brain tumor segmentation, suggesting its applicability to various clinical scenarios.

One of the significant improvements in our study is the detailed implementation of transfer learning techniques. We employed fine-tuning and ensemble learning using the STAPLE method to enhance the model's performance across diverse datasets, including Adult Glioma, Sub-Saharan Africa (SSA) populations, and Pediatric (PED) brain tumors. Fine-tuning significantly improved the baseline models, with the DSC for Pediatric brain tumors increasing from 0.4097 to 0.6248 and the HD95 decreasing from 163.37 to 37.46. Similarly, for the SSA datasets, the DSC improved from 0.7832 to 0.8647, and the HD95 decreased from 18.38 to 10.98. These results underscore the effectiveness of our transfer learning approach in enhancing segmentation accuracy and generalization.

However, it is important to acknowledge the model's limitations. Despite outperforming previous SOTA methods, there is a scope for improvement in the HD95 metric, which suggests that the HT-CNNs model's ability to capture the full extent of tumor boundaries can be enhanced. Additionally, the transfer learning approach, while beneficial, depends on the quality and diversity of the source datasets; the performance gains might not be as pronounced when transferring knowledge from significantly dissimilar domains.

Furthermore, the qualitative analysis of segmentation predictions underscores the practical effectiveness of the HT-CNNs model. The ability to produce high-quality segmentations across different datasets confirms the model's robustness and underscores its potential for real-world clinical applications. The precision with which HT-CNNs delineates tumor boundaries can have a direct impact on treatment planning and patient outcomes, emphasizing the clinical relevance of our research.

## Conclusion

HT-CNNs sets a new standard for brain tumor segmentation, leveraging an ensemble approach and the versatility of transfer learning. This method has proven effective in enhancing accuracy and generalization across diverse segmentation tasks, even with limited data. The fine-tuning and ensemble learning techniques, particularly the use of the STAPLE method, have significantly improved segmentation performance, as evidenced by the substantial increases in DSC and decreases in HD95 metrics. The publicly available code and model docker image encourage further research and adaptation in clinical contexts.

The success of HT-CNNs invites future work to explore the refinement of attention mechanisms to improve the capture of tumor boundaries and the generalization to other types of medical imaging tasks. Limitations such as computational demands and the need for more extensive validation in clinical settings are areas we aim to address moving forward. The true test of HT-CNNs will be its performance in real-world clinical applications, where it has the potential to aid in treatment planning and improve patient outcomes. Acknowledging these limitations is essential as we continue to iterate on our model, ensuring it can meet the rigorous demands of clinical use and ultimately enhance patient care for those suffering from brain tumors.